\documentstyle[prl,aps,epsf,floats]{revtex}

\begin{document}
\draft

\twocolumn[\hsize\textwidth\columnwidth\hsize\csname @twocolumnfalse\endcsname

\title{
Bi-stable tunneling current through a molecular quantum dot
}
\author{
A.S. Alexandrov$^{1,2}$,  A.M.  Bratkovsky$^1$, and 
R. Stanley Williams$^1$
}
\address{$^1$Hewlett-Packard Laboratories, 1501 Page Mill 
Road, 1L, Palo Alto, California 94304\\
$^2$Department of Physics, Loughborough University,
Loughborough LE11 3TU, United Kingdom}
\date{April 17, 2002}
\maketitle

\begin{abstract}
An exact solution is presented for tunneling through a negative-$U$ degenerate molecular 
quantum dot weakly coupled to electrical leads. 
The tunnel current exhibits hysteresis if the level degeneracy 
of the negative-$U$ dot is larger than two. Switching occurs in the voltage range $ V_1 < V < V_2 $
as a result of attractive electron  correlations in
the molecule, which open up a new conducting channel when the voltage is above the threshold bias voltage $V_2$. Once this current has been established, 
the extra channel remains open as the voltage is reduced  down to the lower
threshold voltage $V_1$.   Possible realizations of bi-stable molecular quantum
 dots are fullerenes, especially $C_{60}$, and mixed-valence compounds.

\end{abstract}

\vskip2pc]

\narrowtext

Molecular-scale electronics is currently a very active area of research\cite
{mark98}. The present goals for this field are to design and characterize
molecules that could be the ``transmission lines'' \cite{lehn90,tour00} and
active elements in electronic circuitry\cite{mark98,pat99}. The dominant
mechanism of transport through active devices will most likely be resonant
tunneling through electronic molecular states \cite{pat99}(see also \cite
{restunn,kb01} and references therein). A few experimental studies \cite
{pat99,exp} provide evidence for various molecular switching effects, where
the current-voltage (I-V) characteristics show two branches with high and
low current for the same voltage. This remarkable phenomenon can result from
a conformational transformation of certain molecules containing a ``moving
part'' like a bypyridinium ring, which changes its position if the voltage
is sufficiently high. The transformation necessarily involves a large
displacement of many atoms so that this $ionic$ switching is rather slow,
perhaps operating on a millisecond scale. Switching has also been observed
for simple molecules (organic molecular films), and in some cases is
strongly dependent on the choice of contacts and substrates\cite{exp}.
Molecular devices that exhibit bi-stability and fast switching could be the
basis of future oscillators, amplifiers and other important circuit
elements. Thus, further progress in molecular electronics will depend upon
finding molecules and understanding {\em intrinsic} mechanisms for their
reversible switching from low- to high-current states.

In this Letter, we study a model quantum dot, which exhibits an intrinsic $%
electronic$ \ switching \ of the current state due to {\em attractive}
electron correlations. We show that if the degeneracy is larger than two,
the tunnel current becomes bistable in some voltage range and the dot
exhibits a current hysteresis as a function of bias voltage. In the simplest
case of a doubly degenerate level, the bistability does {\em not} occur. We
present the exact solution of the model, allowing for a detailed analysis of
the current bistability.

{\em Repulsive} electron correlations cause the ``Coulomb blockade'' in the
I-V characteristics of quantum dots \cite{win}. However, they cannot cause
any switching. Here, we show that a negative Hubbard $U$ of any origin can
provide an intrinsic non-retarded current switching of a molecular quantum
dot. One mechanism, that can produce a negative $U$ in molecular systems, is a
strong electron-phonon (vibronic) interaction. If the tunneling time is
comparable to or larger than the characteristic phonon times, a polaron is
formed inside the molecular wire \cite{fis}. There is a wide range of bulk
molecular conductors with polaronic carriers. Since the formation of
polarons in polyacetylene (PA) was theoretically discussed \cite{su80}, they
were detected optically in PA \cite{feldblum82}, in conjugated polymers such
as polyphenylene, polypyrrole, polythiophene, polyphenylene sulfide \cite
{bredas84}, Cs-doped biphenyl \cite{ramsey90}, n-doped bithiophene \cite
{steinmuller93}, polyphenylenevinylene(PPV)-based light emitting diodes \cite
{swanson93}, and other molecular systems. In contrast to bare electrons,
polarons attract each other at short distances of the order of the
interatomic spacing and form small bipolarons \cite{alemot}. Bipolaron
formation can strongly affect the transport properties of long molecular
wires, as discussed recently \cite{abk}. When bipolarons are not formed in
molecular quantum dots because of the short life-time of the carrier inside
the molecule, the attractive correlations between carriers still remain.
Moreover, attractive short-range correlations (negative Hubbard $U$) are
feasible even without electron-phonon interactions. For example, they might
be of a pure ``chemical'' origin, as in the mixed valence complexes \cite
{chemistry}.

Our starting point is the tunneling Hamiltonian, which includes a negative
Hubbard $U$ in the molecular eigenstate $\varepsilon _{\mu }$ coupled with
the $left$ and $right$ leads by the hopping integrals $h _{\alpha k\mu }
$ 
\begin{eqnarray}
H &=&\sum_{_{\mu }}\varepsilon _{_{\mu }}\hat{n}_{_{\mu }}+\frac{1}{2}%
U\sum_{_{\mu }\neq \mu ^{\prime }}\hat{n}_{_{\mu }}\hat{n}_{\mu ^{\prime
}}+\sum_{k,\alpha }\xi _{\alpha k}a_{\alpha k}^{\dagger }a_{\alpha
k}\nonumber \\ 
&&+\sum_{k,\mu ,\alpha }(h _{\alpha k\mu }a_{\alpha k}^{\dagger }c_{\mu
}+H.c.).
\label{eq:ham}
\end{eqnarray}
Here $a_{\alpha k}$ and $c_{_{\mu }}$ are the annihilation operators in the
left ($\alpha =1)$ and right ($\alpha =2)$ leads, and in the molecule,
respectively, $\hat{n}_{\mu }=c_{\mu }^{\dagger }c_{\mu }$, $\xi _{\alpha k}$
is the energy dispersion in the leads, and $U<0.$ This negative $U$
Hamiltonian is derived from the general microscopic Hamiltonian with the
bare electron-phonon (vibron) interaction and Coulomb repulsion using the
canonical transformation\cite{alemot}. It was successfully applied to glassy
semiconductors \cite{and,mot}, high $T_{c}$ superconductors \cite{alemot,mic}
including doped fullerenes \cite{kab} mixed valence compounds\cite{chemistry}%
, and more recently to superconductor-insulator-superconductor tunneling 
\cite{kiv}. The current through the molecular quantum dot is conveniently
expressed in terms of the molecular density of states (DOS) $\rho _{\mu
}(\omega )$ as \cite{win} 
\begin{equation}
I=e\int_{-\infty }^{\infty }d\omega \left[ f_{1}(\omega )-f_{2}(\omega )%
\right] \sum_{\mu }\Gamma _\mu (\omega )\rho _{\mu }(\omega ),
\label{eq:current}
\end{equation}
where $f_{1,2}(\omega )=\{\exp [(\omega +\Delta \mp eV/2)/T]+1\}^{-1},$ $%
\Delta $ is the position of the lowest unoccupied molecular level with
respect to the chemical potential, $\Gamma _\mu(\omega )=\Gamma
_{1\mu}(\omega )\Gamma _{2\mu }(\omega )/[\Gamma _{1\mu }(\omega
))+\Gamma _{2\mu }(\omega )]$, and $\Gamma _{\alpha \mu }(\omega
)=2\pi \sum_{\alpha k }|h _{\alpha k\mu }|^{2}\delta (\omega -\xi
_{\alpha k}).$ Molecular DOS is given by $\rho _{\mu }(\omega )=-(1/\pi )%
\mathop{\rm Im}%
G_{\mu }^{(1)}(\omega ),$ where $G_{\mu }^{(r)}(\omega )$ is the Fourier
transform of the $r$-particle retarded Green's function (GF) defined as 
\begin{equation}
G_{\mu }^{(r)}(t)=-i\theta (t)\sum_{\mu _{1}\neq \mu _{2}\neq ...\mu
}\left\langle \left\{ c_{\mu }(t)\prod_{i=1}^{r-1}\hat{n}_{\mu
_{i}}(t),c_{\mu }^{\dagger }\right\} \right\rangle 
\end{equation}
for $2\leq r<\infty $ and $G_{\mu }^{(1)}(t)=-i\theta (t)\left\langle
\left\{ c_{\mu }(t),c_{\mu }^{\dagger }\right\} \right\rangle .$ Here $%
\left\{ {\dots,\dots }\right\} $ is the anticommutator, and $\theta (t)=1$ for $t>0$ and
zero otherwise. In the following, we apply perturbation theory with respect
to the hopping integrals, neglecting any contribution to the current other
than $h ^{2}$, but keeping all orders of the negative Hubbard $U.$
Terms of higher order in $h $ cannot change the gross I-V features for
any voltage except the narrow transition regime from one lead to another.
Then, applying the equations of motion for the Heisenberg operators $c_{\mu
}(t),\hat{n}_{\mu }(t)$ and $a_{\alpha k}(t)$, we obtain an {\it infinite}
set of coupled equations for the molecular GFs as 
\begin{eqnarray}
i\frac{dG_{\mu }^{(r)}(t)}{dt} &=&\delta (t)\sum_{\mu _{1}\neq \mu _{2}\neq
...\mu }\prod_{i=1}^{r-1}n_{\mu _{i}}(0)+  \nonumber \\
&&[\varepsilon _{_{\mu }}+(r-1)U]G_{\mu }^{(r)}(t)+UG_{\mu }^{(r+1)}(t),
\end{eqnarray}
where $n_{\mu }(t)=\left\langle c_{\mu }^{\dagger }(t)c_{\mu
}(t)\right\rangle $ is the expectation number of electrons on the conducting
molecular level. For the sake of analytical transparency we solve this
system for a molecule having one $d$-fold degenerate energy level with $%
\varepsilon _{_{\mu }}=0.$ In this case the set appears to be finite, and it
can be solved by using the Fourier transformation. Fourier transforming Eqs.(4)
we find the one-particle GF as 
\begin{equation}
G_{\mu }^{(1)}(\omega )=\sum_{r=0}^{d-1}\frac{Z_{r}(n)}{\omega -rU+i\delta },
\label{eq:G1}
\end{equation}
where $\delta =+0$, $n=n_{\mu }(0),$ and 
\[
Z_{r}(n)=\frac{(d-1)!}{r!(d-1-r)!}n^{r}(1-n)^{d-1-r}.
\]
This is an exact solution with respect to correlations which satisfies all
sum rules. The electron density $n_{\mu }(t)$ obeys the rate equation, which
is obtained by using the equations of motion as 
\begin{equation}
\frac{dn_{\mu }(t)}{dt}=2\sum_{\alpha ,k}h _{\alpha k\mu }%
\mathop{\rm Im}%
A_{\alpha k\mu }^{(1)}(t),
\label{eq:motion}
\end{equation}
where 
\begin{equation}
A_{\alpha k\mu }^{(r)}(t)=\sum_{\mu _{1}\neq \mu _{2}\neq ...\mu
}\left\langle c_{\mu }^{\dagger }(t)\prod_{i=1}^{r-1}\hat{n}_{\mu
_{i}}(t)a_{\alpha k}(t)\right\rangle .
\end{equation}
These correlation functions should be calculated to first order with respect
to the hopping integrals $h $. In this order, they satisfy the infinite
set of coupled equations 
\begin{eqnarray}
i\frac{dA_{\alpha k\mu }^{(r)}(t)}{dt} &=&h _{\alpha k\mu }[n_{\mu
}(t)-f(\xi _{\alpha k})]\sum_{\mu _{2}\neq ...\mu }\prod_{i=1}^{r-1}n_{\mu
_{i}}(t)+  \nonumber \\
&&UA_{\alpha k\mu }^{(r+1)}(t)+[\xi _{\alpha k}-(r-1)U]A_{\alpha k\mu
}^{(r)}(t).
\end{eqnarray}
But if we have a finite number of molecular states, the set is finite like
Eqs.(4). One readily solves this set in the steady state, when $n_{\mu
_{i}}(t)$ and $A_{\alpha k\mu }^{(r)}(t)$ become time-independent. For a $d$%
-fold degenerate energy level, \ the one-particle correlation function is
found as 
\begin{equation}
A_{\alpha k\mu }^{(1)}=[n-f(\xi _{\alpha k})]h _{\alpha k\mu
}\sum_{r=0}^{d-1}\frac{Z_{r}(n)}{\xi _{\alpha k}-rU+i\delta }.
\label{eq:Aop}
\end{equation}
Substituting Eq.~(\ref{eq:Aop}) into Eq.~(\ref{eq:motion}), we obtain
the steady state equation for 
the electron density on the molecule as 
\begin{equation}
\sum_{\alpha }\sum_{r=0}^{d-1}\Gamma _{\alpha }(rU)[n-f_{\alpha
}(rU)]Z_{r}(n)=0.
\end{equation}
Here we assume that $\Gamma _{\alpha \mu }(\omega )=$ $\Gamma _{\alpha
}(\omega )$ does not depend on $\mu ,$ otherwise the degeneracy would be
removed. To simplify the mathematics further, we now assume that $\Gamma
_\alpha (\omega )=\Gamma $ is a constant. Then, from Eq.~(\ref{eq:current}) and
Eq.~(\ref{eq:G1}) the current is found as 
\begin{equation}
j=\sum_{r=0}^{d-1}[f_{1}(rU)-f_{2}(rU)]Z_{r}(n),
\end{equation}
where $j=I/I_{0}$ with $I_{0}=ed\Gamma /2$. Let us consider two-fold,
four-fold, and six-fold degenerate molecular level. For $d=2$ the kinetic
equation is linear in $n$, and there is only one solution, 
\begin{equation}
n=\frac{\sum_{\alpha }f_{\alpha }(0)}{2+\sum_{\alpha }[f_{\alpha
}(0)-f_{\alpha }(U)]}.
\end{equation}
The current through a two-fold degenerate molecular dot is found as 
\begin{equation}
j=2\frac{f_{1}(0)[1-f_{2}(U)]-f_{2}(0)[1-f_{1}(U)]}{2+\sum_{\alpha
}[f_{\alpha }(0)-f_{\alpha }(U)]}.
\end{equation}
There is no current bistability in this case. Moreover, if the temperature
is low ($T\ll \Delta ,|U|$) there is practically no effect of correlations
on the current, $j\approx V\theta (e|V|-2\Delta )/|V|.$ 

Remarkably,
four-fold or higher- degenerate negative $U$ dots reveal a switching effect.
In this case, the kinetic equation is nonlinear, allowing for a few
solutions. If $eV<2(\Delta -|U|),$ the only physically allowed solution of
Eq.(10) for $d=4$ and $d=6$ at zero temperature is $n=0.$ If $2(\Delta
-|U|)<eV<2\Delta $, $T=0$ and $|U|<2\Delta /d,$ the kinetic equation is
reduced to 
\begin{equation}
2n=1-(1-n)^{d-1}.
\label{eq:nd}
\end{equation}
For $d=4$ it has $two$\ physical roots, $n=0$ and $n=(3-5^{1/2})/2\approx
0.38$ \cite{erm}$.$ In this voltage range $f_{1}(0)=f_{2}(rU)=0,$ but $%
f_{1}(U)=f_{1}(2U)=f_{1}(3U)=1$ at $T=0,$ when $|U|<\Delta /2.$ Using the
sum rule $\sum_{r=0}^{d-1}Z_{r}(n)=1$ and the kinetic equation (10), the
current is simplified in this voltage range as $j=2n.$ Hence we obtain two
stationary states of the molecule with low (zero at $T=0$) and high current, 
$I\approx 0.76I_{0}$ for the same voltage in the range $2(\Delta
-|U|)<eV<2\Delta .$ For $d=6$, the kinetic equation has two physical roots
in this voltage range, $n=0$ and $n\approx 0.48,$ which corresponds to $I=0$
and $I\approx 0.96I_{0},$ respectively. Above the standard threshold, $%
eV>2\Delta ,$ where $f_{1}(rU)=1$ and $f_{2}(rU)=0,$ when $|U|<\Delta /3,$
there is only one solution, $n=0.5$ with the current $I=I_{0}.$

One can better understand the origin of the switching phenomenon by taking
the limit $d\gg 1.$ The physical roots of Eq.~(\ref{eq:nd}) are $n=0$ and $n=0.5$ in
this limit with the current $I=0$ and $I=I_{0}$, respectively. This is
precisely the solution of the problem in the mean-field approximation (MFA),
which is a reasonable approximation for $d\gg 1.$ Indeed, using MFA one
replaces the exact two-body interaction in the Hamiltonian for a mean-field
potential as $\frac{1}{2}U\sum_{_{\mu }\neq \mu ^{\prime }}\hat{n}_{_{\mu }}%
\hat{n}_{\mu ^{\prime }}\approx U\sum_{_{\mu }\neq \mu ^{\prime }}\hat{n}%
_{_{\mu }}n_{\mu ^{\prime }}-\frac{1}{2}U\sum_{_{\mu }\neq \mu ^{\prime
}}n_{_{\mu }}n_{\mu ^{\prime }}$. Then the MFA DOS is given by $\rho _{\mu
}(\omega )=\delta \lbrack \omega -U(d-1)n].$ Using the Fermi-Dirac Golden
rule the rate equation for $n$ becomes 
\begin{equation}
\frac{dn}{dt}=-2\Gamma n+\Gamma \sum_{\alpha }f_{\alpha }[nU(d-1)].
\label{eq:mfa}
\end{equation}
For $T=0$ there are two stationary solutions of Eq.~(\ref{eq:mfa}), $n=0$ and $n=0.5$
in the voltage range $2(\Delta -|\tilde{U}|)<eV<2\Delta $, and only one
solution, $n=0.5$ for $eV>2\Delta ,$ where $\tilde{U}=U(d-1)/2.$ The MFA
current is found as 
\begin{equation}
j=f_{1}(2n\tilde{U})-f_{2}(2n\tilde{U}).
\end{equation}
Combining this equation and the rate equation (15) with $dn/dt=0$ , we
obtain the I-V characteristic equation as 
\begin{eqnarray}
&&\frac{|\tilde{U}|}{\Delta }(1-R) = 1-\frac{T}{\Delta } \nonumber\\
&&\times\ln \left[ \frac{(1+R)\sinh \left( eV/2T\right) }{j}-\cosh \left(
eV/2T\right) \right],
\end{eqnarray}
where 
\begin{equation}
R=\left\{ \left[ 1-j\coth (eV/2T)\right] ^{1/2}-\frac{j^{2}}{\sinh
^{2}(eV/2T)}\right\} ^{1/2}.
\end{equation}
The $I/V$ curves are shown in Fig.1 for different temperatures and $|\tilde{U%
}|=0.9\Delta .$ Interestingly, the temperature narrows the voltage range of
the hysteresis loop, but the transition from the low (high)-current branch
to the high (low)-current branch remains discontinuous. 
\begin{figure}[t]
\epsfxsize=3.8in \epsffile{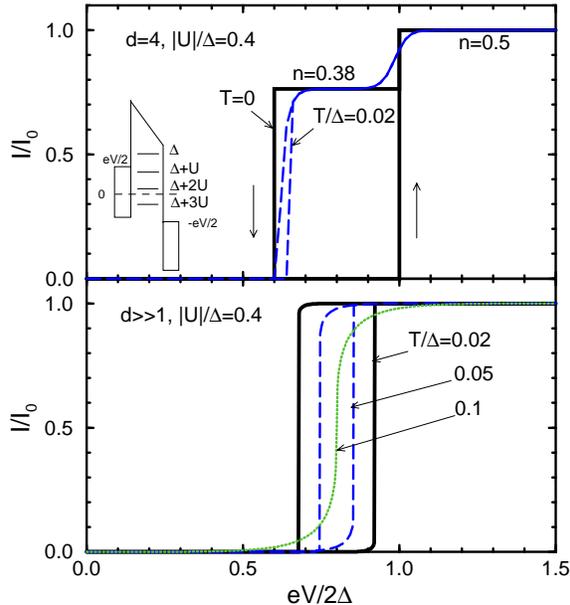}
\caption{
 The current-voltage hysteresis loop for the negative-$U$
 molecular quantum dot for the molecular level with the degeneracies
 $d=4$ and $d \gg 1$. The two cases are very different in some
 aspects. Top panel: $d=4$, $|U|/\Delta=0.4$. In the hysteretic region
 the current jumps to the value $I/I_0=0.76$ at the lower threshold
 ($eV_1/2\Delta=0.6$) and then to the value $I/I_0=1$ at the higher
 threshold ($eV_2/2\Delta=1$). Bottom panel: highly degenerate
 molecular level with $d\gg 1$, $|\tilde U|/\Delta=0.4$. The current jumps
 directly to the value $I/I_0=1$ at the lower threshold. Temperature
 quickly reduces the width of the hysteretic region, much faster for
 $d=4$ than for $d \gg 1$. In the latter case the hysteresis is over
 at about $T/\Delta=0.1$. Inset in the top panel shows a schematic of
 the energy level diagram for $d=4$ quantum dot under bias voltage
 $V$. 
}
\label{fig:fringe}
\end{figure}

Let us examine the stability of each branch in the framework of the MFA rate
equation (\ref{eq:mfa}). Introducing small fluctuations of the electron density as $%
n(t)=n+\delta n\exp (\gamma t)$ and linearizing Eq.(15) with respect to $%
\delta n$ we find the increment $\gamma ,$%
\begin{eqnarray}
\gamma &=&-2\Gamma +\frac{|\tilde{U}|\Gamma }{2T}\cosh ^{-2}\left( \frac{%
\Delta -2n|\tilde{U}|-eV/2}{2T}\right)  \nonumber \\
&&+\frac{|\tilde{U}|\Gamma }{2T}\cosh ^{-2}\left( \frac{\Delta -2n|\tilde{U}%
|+eV/2}{2T}\right) .
\end{eqnarray}
One can see from this equation that at temperatures $T\ll |\tilde{U}|$ the
low-current branch ($n=0)$ looses its stability at the threshold $%
V_{2}=2\Delta /e,$ while the high-current branch looses its stability at $%
V_{1}=2$ $(\Delta -|\tilde{U}|)/e.$ In the voltage range $V_{1}<V<V_{2}$
both branches are stable, $\gamma \approx -2\Gamma <0.$

Finally, let us analyze the effect of a splitting of the degenerate
molecular level on the bi-stability. The degeneracy could be removed because
of Jahn-Teller distortions and/or the coupling with the leads. We assume
that $d\gg 1$ levels are evenly distributed in a band of a width $W$. Then
the MFA rate equation (15) is modified as 
\begin{equation}
\frac{dn_{\mu }}{dt}=-2\Gamma n_{\mu }+\Gamma \sum_{\alpha }f_{\alpha
}(\varepsilon _{_{\mu }}+NU),
\end{equation}
where $N=\sum_{\mu ^{\prime }}n_{\mu ^{\prime }}.$ For $T=0$ in the
stationary regime ($dn_{\mu }/dt=0)$ it has two solutions, $N=0$ and $N=d/2,$
in the voltage range $V_{1}+W<V<V_{2}$ with the current $j=0$ and $j=1$,
respectively. We conclude that the level splitting $W<|U|$ leads to a
narrowing of the voltage range of the bi-stability similar to the
temperature narrowing shown in Fig.1. A parameter-free estimate of the
negative Hubbard $U$ in some oxides yields $|U|$ about a few tens of eV \cite
{alebra}. We expect a negative $U$ of the same order of magnitude in
carbon-based compounds. Among potential candidates for the negative $U$
quantum dot are a single $C_{60}$ molecule ($d=6$), where the
electron-vibronic coupling proved to be particularly strong \cite{par}, or
other fullerenes including short nanotubes ($d\gg 1$) connected to metal
electrodes. Other likely candidates are mixed-valence molecular complexes 
\cite{chemistry}. There should be no retardation of the switching on the
time scale above the inverse vibron (phonon) frequency, which is 10$^{-14}$s
or less in carbon-based compounds.

In conclusion, we have introduced and solved a model for tunneling through a
negative-$U$ degenerate molecular dot weakly coupled to electrical leads.
The exact many-particle Green's functions of a $d-$fold degenerate molecular
level, the density of states and the nonlinear rate equation for the
electron density on the molecule have been derived. We have found the exact
solutions for the carrier population in the dot and the current for 
the degeneracies of the molecular level
$d=2,4,6, $ and $d\gg 1$. The current-voltage characteristics show a
hysteretic behavior for $d>2$ over a finite voltage range. When the voltage
increases from zero, the molecule remains in a low-current state until the
threshold $V_{2}$ is reached. Remarkably, when the voltage {\em decreases}
from the value above the threshold $2\Delta /e$, the molecule remains in the
high-current state down to the voltage $V_{1}=$($2\Delta -|U|)/e$, well
below the threshold $V_{2}$. This mechanism for electronic molecular
switching without retardation requires many-particle attractive
correlations, which can arise from strong electron-vibronic coupling and/or
mixed valence states. Experimental verification of such bi-stable systems
will require careful collection and analysis of both forward and reverse
voltage sweeps of the tunneling current through candidate molecules. The
forward voltage sweep by itself will resemble a standard Coulomb blockade
I-V characteristic with a turn-on voltage of $V_{2}$, whereas the reverse
sweep should reveal hysteresis.

This work has been partly supported by DARPA. The authors acknowledge useful
discussions with P. Kornilovich and D. Stewart.

\end{document}